\documentclass[prd,groupedaddress,showpacs,showkeys,onecolumn,nofootinbib]{revtex4}
\usepackage{amsfonts,amssymb,amsmath,amsthm,amscd}
\usepackage{graphicx}
\usepackage{dcolumn}
\usepackage{hyperref}
\usepackage{bm}
\usepackage{enumerate}
\begin{document}
\title{On the emergence of Lorentzian signature and scalar gravity}
\author{F. Girelli}
\email{girelli@sissa.it}
\author{S. Liberati}
\email{liberati@sissa.it}
\author{L. Sindoni}
\email{sindoni@sissa.it}
\affiliation{SISSA, Via Beirut 2-4, 34014 Trieste, Italy and INFN sezione di Trieste}
\date{\today}
\bigskip
\begin{abstract}
\bigskip

In recent years, a growing momentum  has been gained by the emergent gravity framework. Within the latter, the very concepts of geometry and gravitational interaction are not seen as elementary aspects of Nature but rather as collective phenomena associated to the dynamics of more fundamental objects.  In this paper we want to further explore this possibility by proposing a model of emergent Lorentzian signature and scalar gravity. Assuming that the dynamics of the fundamental objects can give rise in first place to a Riemannian manifold and a set of scalar fields we show how time (in the sense of hyperbolic equations) can emerge as a property of perturbations dynamics around some specific class of solutions of the field equations. Moreover, we show that these perturbations can give rise to a spin-0 gravity via a suitable redefinition of the fields that identifies the relevant degrees of freedom. In particular, we find that our model gives rise to Nordstr\"om gravity. Since this theory is invariant under general coordinate transformations, this also shows how diffeomorphism invariance (albeit of a weaker form than the one of general relativity) can emerge from much simpler systems.

\end{abstract}
\maketitle
\def\wt{\widetilde}
\def\gsim{\; \raisebox{-.8ex}{$\stackrel{\textstyle >}{\sim}$}\;}
\def\lsim{\; \raisebox{-.8ex}{$\stackrel{\textstyle <}{\sim}$}\;}
\def\half{{1\over2}}
\def\a{\alpha}
\def\b{\beta}
\def\g{\gamma}
\def\d{\delta}
\def\e{\epsilon}
\def\o{\omega}
\def\m{\mu}
\def\t{\tau}
\def\L{{\mathcal L}}
\def\p{{\mathbf{p}}}
\def\q{{\mathbf{q}}}
\def\k{{\mathbf{k}}}
\def\fp{{p_{\rm 4}}}
\def\fq{{q_{\rm 4}}}
\def\fk{{k_{\rm 4}}}
\def\etal{{\emph{et al}}}
\def\det{{\mathrm{det}}}
\def\tr{{\mathrm{tr}}}
\def\aka{{\emph{aka}}}
\def\b{$\blacktriangleright$ }
\def\demi{\frac{1}{2} } \def\C{\mathbb{C}}
\def\P{\mathbb{P}} \newcommand{\dr}{\rightarrow} \newcommand{\N}{\mathbb{N}} \newcommand{\Z}{\mathbb{Z}} \newcommand{\Q}{\mathbb{Q}}
\newcommand{\R}{\mathbb{R}} \newcommand{\K}{\mathbb{K}} \newcommand{\He}{\mathbb{H}} \newcommand{\cs}{C${}^*$}

\def\aa{{\cal A}} \def\bb{{\cal B}} \def\cc{{\cal C}} \def\dd{{\cal D}} \def\Ee{{\cal E}} \def\ff{{\cal F}} 
\def\hh{{\cal H}} \def\ii{{\cal I}} \def\jj{{\cal J}} \def\kk{{\cal K}} \def\lll{{\cal L}} \def\mm{{\cal M}} \def\nn{{\nonumber}}
\def\oo{{\cal O}} \def\ppp{{\cal P}} \def\qq{{\cal Q}} \def\rr{{\cal R}} \def\ss{{\cal S}} \def\tt{{\cal T}} \def\uu{{\cal U}}
\def\vv{{\cal V}} \def\ww{{\cal W}} \def\xx{{\cal X}} \def\yy{{\cal Y}} \def\zz{{\cal Z}}

\def\bl{{\mbox{\boldmath $l$}}}
\def\bv{{\mbox{\boldmath $v$}}}
\def\bw{{\mbox{\boldmath $w$}}}
\def\ba{{\mbox{\boldmath $a$}}}
\def\bx{{\mbox{\boldmath $x$}}}
\def\bX{{\mbox{\boldmath $X$}}}
\def\bs{{\mbox{\boldmath $s$}}}
\def\by{{\mbox{\boldmath $y$}}}
\def\bj{{\mbox{\boldmath $j$}}}
\def\bz{{\mbox{\boldmath $z$}}}
\def\bp{{\mbox{\boldmath $p$}}}
\def\bP{{\mbox{\boldmath $P$}}}
\def\bq{{\mbox{\boldmath $q$}}}
\def\bd{{\mbox{\boldmath $d$}}}
\def\bJ{{\mbox{\boldmath $J$}}}
\def\bS{{\mbox{\boldmath $S$}}}
\def\bP{{\mbox{\boldmath $P$}}}
\def\bn{{\mbox{\boldmath $n$}}}
\def\poi#1{\{ #1 \}}
\def\com#1{[ #1 ]}
\def\ka{{\kappa}} \newcommand{\mone}{^{-1}}
\newcommand{\be}{\begin{equation}}
\newcommand{\bee}{\begin{equation}}
\newcommand{\eeq}{\end{equation}} 
\def\ot{{\otimes}}
\def\g{\mathfrak{g}}
\def\mmm{\mathfrak{m}}
\def\nnn{\mathfrak{n}}
\def\cop{\Delta} \def\odel{\overline\delta}
\def\ov{\overline}
\newcommand{\ket}[1]{|#1\rangle} \newcommand{\bra}[1]{\langle #1|} \newcommand{\bk}[1]{\langle #1 \rangle}
\def\la{\langle} \def\ra{\rangle} \newcommand{\one}{\mbox{$1 \hspace{-1.0mm}  {\bf l}$}}
\def\f{\frac} \def\t{\textrm{Tr}} \def\MP{M_P}
\def\dag{^\dagger}
\newcommand{\ovl}[1]{\overline{#1}} \newcommand{\bem}{\begin{bmatrix}} \newcommand{\eem}{\end{bmatrix}}
\newcommand{\beq}{\begin{equation}} \newcommand{\ee}{\end{equation}} \newcommand{\eq}{\end{equation}}
\newcommand{\beqa}{\begin{eqnarray}} \newcommand{\eeqa}{\end{eqnarray}}
\def\ka{\kappa}
\def\ie{{i.e. \/}}
\def\eg{{e.g. \/}}
\def\cf{{c.f. \/}}
\def\dx{{\bf [dx^d]}}
\def\meas{{\bf [d\mu]}}
\def\x{{\bf x}}
\def\mn{{\mu\nu}}
\def\mmm{{\mathfrak{m}}}
\def\hphi{\hat\phi}
\def\htheta{\hat\theta_1}
\def\hn{\hat n_1}

\def\HRULE{{\bigskip\hrule\bigskip}}
\section{Introduction}
In spite of being the first force of Nature to be understood in physical terms, gravity is somehow still a riddle for  physicists. Not only it keeps evading a full quantum description as well as any form of unification with the other interactions, it also puzzles us with profound questions and unexpected features.
We will not attempt here to present a complete list of these startling aspects of gravitation theory, but we can recall, for example, the surprising connection between gravity and thermodynamics associated to black hole physics \cite{Bardeen:1973gs,Hawking:1974sw,Jacobson} as well as the deep questions associated to the nature of inertia and time \cite{Isham,Kuchar,Barbour}.

In recent years, a new approach to these old problems has been gaining momentum and many authors have been advancing the idea that gravity could all in all be an intrinsically classic/large scale phenomenon similar to a condensed matter state made of many atoms~\cite{BLH}. In this sense gravity would not be a fundamental interaction but rather a large scale/numbers effect, something emergent from a quite different dynamics of some elementary quantum objects. In this sense, many examples can be brought up, starting from the causal set proposal \cite{Bombelli:1987aa}, passing to group field theory \cite{Oriti:2007qd} or the recent quantum graphity models \cite{Konopka:2008hp} and other approaches (see \eg \cite{Dreyer}).

All these models and many others share a common scheme:  they consider a fundamental theory which is not General Relativity and examine, using different techniques often borrowed from condensed matter physics, how space, time and their dynamics could emerge in some regime. In this sense a leading inspirational role also been played by another stream of research which goes under the name of ``analogue models of gravity" \cite{LRR}. These are condensed matter systems which have provided toy models showing how at least the concept of a pseudo-Riemannian metric and Lorentz invariance of matter equations of motion can be emergent. For example, non-relativistic systems which admit some hydrodynamics description can be shown to have perturbations (phonons) whose propagation is described, at low energies, by hyperbolic wave equations on an effective Lorentzian geometry~\cite{LRR}. While these models have not provided so far  also an analogy of emergent gravitational dynamics equations they do have provided a new stream of ideas about many other pressing problems in gravitation theory (see for example recent works on the origin of the cosmological constant in emergent gravity~\cite{L-Vol}).

In this paper we shall try to further advance in the direction of a consistent emergent gravity picture by proposing a toy model that will show some of the salient features that would be desirable to  see in an eventual full theory of emergent gravity. In particular we shall focus here on the possible dynamical origin of time\footnote{For early, alternative work in this direction see also \cite{greensite} and references therein.} (in the sense of pseudo-Riemannian signature of the metric) as well as of another crucial aspect of gravitation theory, \ie diffeomorphism invariance (see however the discussion in the conclusion).
It is perhaps important to stress here, that a Lorentzian signature has been considered by some authors as a necessary, but not sufficient, condition for the emergence of time. This is because these authors associate to the latter not just a special signature, but also to a dimension with a univocal direction (arrow of time), which is of course much less easy to rigorously characterize~\cite{Ellis:2006sq}. Missing a general agreement on this issue, we will stick here to the more technical language of Lorentzian signature, although we think that this feature is the very essence of the nature of time.

In building up our model we shall make the conjecture that the dynamics of some elementary quantum objects (what one could call the ``atoms of spacetime") is such to produce, in some semi-classical or large number limit  (hydrodynamic limit, or  condensation...), a Riemannian manifold (essentially $\R^4$ equipped with the trivial metric $\delta_\mn$), a set of scalar fields $\Psi_i$, and their  Lagrangian $\lll$. Hence we do not assume a priori any notion of time. Instead we shall consider, largely following the intuition gained from condensed matter analogues of gravity, the possibility that Lorentzian dynamics can emerge as a property of the equations associated to perturbations around some solutions of the equations of motion. These perturbations will be associated to the emergent gravity and matter degrees of freedom which will characterize our gravitational dynamics.

The plan of the paper will be the following. In the first section, we shall look for a specific Lagrangian  so that the perturbations around some solutions of the equations of motion will see an effective pseudo-Riemannian metric, even though the fundamental theory is Euclidean. In other words, we shall be looking for specific properties of the Lagrangian so that  time can  emerge.

In the second section, we shall consider the perturbations around a specific solution of the equations of motion (\eg such that these perturbations propagate on Minkowski spacetime), and identify among them the degrees of freedom corresponding to matter and to gravity. More precisely, we shall show how the equations of motion for the perturbations can be reinterpreted as the Einstein--Fokker equations of motion. These latter are historically the first  equations of motion for a relativistic theory of gravity, written in a diffeomorphism invariant way. They describe  a relativistic scalar gravity theory known as Nordstr\"om gravity. Our construction identifies therefore a possible mechanism to obtain diffeomorphisms invariant equations of motion for gravity, in an emergent framework.

\section{Time emergence}

As explained in the Introduction, we consider that the fundamental unknown theory gives rise in some large number limit to  simple structures such as $\R^4$ equipped with the Euclidean metric $\delta^\mn$, and a set of scalar fields $\Psi_i(x_\mu)$, $i=1,...,N$ ($x_\mu\in\R^4$) with their Euclidean Lagrangian $\lll$. Since we do not know this fundamental theory, we choose such Lagrangian to be of the simple shape\footnote{We could also consider  a dependence on crossed terms of the kind $h^{\mn}\partial_\mu\Psi_i\partial_\nu\Psi_j$, however this is not changing the final result.}
\begin{equation}\label{fundamental Lagrangian}
\lll=F(X_1,..., X_N).
\end{equation}
with $X_i=\delta^{\mn}\partial_\mu\Psi_i\partial_\nu\Psi_i$.
It is easy to see that this Lagrangian is invariant under the Euclidean group $ISO(4)$.
The equations of motion are then simply for a given field $\Psi_i$ 
\begin{equation}\label{fundamental EOM}
 \partial_\mu \left(\frac{\partial F}{\partial X_i}\partial^\mu\Psi_i\right)=0=\Sigma_j{\left(\frac{\partial^2F}{\partial X_i\partial X_j}\partial_\mu X_j\right)} \partial^\mu\Psi_i + \frac{\partial F}{\partial X_i}\partial_\mu\partial^\mu\Psi_i.
\end{equation}

Let us now consider a specific solution of the above equations of motion, $\psi_i$ and perturbations $\varphi_i$ around it. For $\Psi_i=\psi_i+\varphi_i$, the kinematic term $X_i$ becomes then
\begin{equation}\label{perturbations}
X_i\dr \ov{X}_i+\delta X_i, \quad \textrm{ with } \quad \ov{X}_i= \delta^{\mn}\partial_\mu\psi_i\partial_\nu\psi_i \quad \textrm{ and } \quad \delta X_i= 2\partial_\mu\psi_i\partial^\mu\varphi_i +\partial_\mu\varphi_i\partial^\mu\varphi_i.
\end{equation}
We intend now to identify some specific $F$ such that the Lagrangian for the  perturbations $\varphi_i$ is invariant under the Poincar\'e group $ISO(3,1)$. To determine the Lagrangian for the perturbations $\varphi_i$, we expand \eqref{fundamental Lagrangian} using \eqref{perturbations}.
\begin{eqnarray}\label{expansion}
F(X_1, .., X_N)\dr && F(\ov{X}_1, ..,\ov{X}_N)+
\sum_j\left.\frac{\partial F}{\partial X_j }\right|_{\ov{X}}\delta X_j \nn \\
&&+\demi \sum_{jk}\left. \frac{\partial^2 F}{\partial X_j\partial X_k }\right|_{\ov{X}} \delta X_j\delta X_k +
 \frac{1}{6}\sum_{jkl}\left.\frac{\partial^3 F}{\partial X_j\partial X_k \partial X_l}\right|_{\ov{X}}\delta X_j\delta X_k\delta X_l+ ...
\end{eqnarray}
The first term $F(\ov{X}_1, .., \ov{X}_N)$ is the Lagrangian for the classical solution $\psi_i$. The second term, the one linear in $\delta X_j$, contains a term linear in $\partial_\mu\varphi_i$, which is zero on shell. We can also identify the quadratic  contribution for $\partial_\mu \varphi_k\partial_\nu \varphi_k$:
\begin{eqnarray}
&& \textrm{for } k\neq l, \quad \partial_\mu \varphi_k \partial_\nu \varphi_l \left(  2 \left. \frac{\partial^2 F}{\partial X_k\partial X_l }\right|_{\ov{X}}\partial^\mu \psi_k\partial^\nu \psi_l\right),\label{crossed terms}\\
&& \textrm{for } k= l, \quad \partial_\mu \varphi_k \partial_\nu \varphi_k \left( \left. \frac{\partial F}{\partial X_k }\right|_{\ov{X}}\delta^{\mn} + \demi \left. \frac{\partial^2 F}{(\partial X_k)^2 }\right|_{\ov{X}}\partial^\mu \psi_k\partial^\nu \psi_k\right) \label{diagonal terms}.
\end{eqnarray}
The contribution \eqref{crossed terms} introduces some mixing between fields in the kinematic term. To simplify the analysis, we demand  that they cancel, which puts a constraint on the choice of $F$, \ie $ \left. \frac{\partial^2 F}{\partial X_k\partial X_l }\right|_{\ov{X}}=0$, if $k\neq l$. A specific solution is then
\begin{equation}\label{first cond for F}
F(X_1, .., X_N)= f_1(X_1)+...+f_N(X_N).
\end{equation}
We can identify in \eqref{diagonal terms} the effective or emergent metrics\footnote{Actually, we show here the inverse metrics from which the actual metrics can be derived once invertibily conditions are imposed. In our case of interest, this will always be true.} for each field $\varphi_k$, (taking into account \eqref{first cond for F})
\begin{equation}
g^\mn _k\equiv \left. \frac{d f_k}{d X_k }\right|_{\ov{X}_k}\delta^{\mn} + \demi \left. \frac{d^2 f_k}{(d X_k)^2 }\right|_{\ov{X}_k}\partial^\mu \psi_k\partial^\nu \psi_k.
\label{eff-metr}
\end{equation}
Since a priori $f_i\neq f_j$ and $\psi_i\neq \psi_j$ if $i\neq j$, we are dealing with a multi-metric structure: each field sees its own metric. However, we can enforce a mono-metric structure
by constraining the solution $\psi_k$ and the derivatives of $f_k$ at $\ov{X}_k$ to be independent of $k$
\begin{equation}\label{second con for F}
f_k=f, \quad \psi_k=\psi, \quad \forall k.
\end{equation}

So far we have just shown that the perturbations around a solution of the field equations on a Riemannian manifold can propagate, for suitably chosen Lagrangians, on an effective geometry which is not the fundamental one, $\delta_{\mu\nu}$, but rather a rank 2 tensor constructed from it and partial derivatives of the chosen background solution.  Note that, in order for this to be possible, it was crucial to have a starting Lagrangian with non-canonical kinetic terms as it can be clearly evinced by the second contribution to the metrics in equation \eqref{eff-metr}.  As a next step, we show now how for some solutions of the equations of motion, such effective metric can be of pseudo-Riemannian form. In fact, we can even ask that the metric (\ref{eff-metr}) is the Minkowski metric $\eta_\mn$. This will put some constraints on the derivative of $f$, evaluated at $\ov{X}=\partial^\mu \psi\partial_\mu \psi$.

In order to do so, we shall need to specify a particular solution, $\bar{\psi}$, of the equations of motion. Let us take it to be an affine function of the coordinates, $\bar{\psi}= \alpha^\mu x_\mu + \beta$. It is easy to check that this is indeed a solution of our field equations \eqref{fundamental EOM}. Moreover, thanks to the $SO(4)$ symmetry, we can always make a rotation such that
\be\label{choice of psi} \bar{\psi}= \alpha x_0 + \beta.\ee
The choice of the coordinate $x_0$ is completely arbitrary, what only matters is that there is one coordinate which is pinpointed.  Finally, we ask for the metric to have the signature $(-,+,+,+)$.  This puts some constraint on the value of  the derivatives of $f$
\begin{eqnarray}
&&\left. \frac{df}{d X }\right|_{\ov{X}} + \demi \left. \frac{d^2 f}{(d X)^2 }\right|_{\ov{X}}\partial^0 \bar{\psi}\partial^0 \bar{\psi}<0,\nn\\
&& \left. \frac{d f}{d X }\right|_{\ov{X}} + \demi \left. \frac{d^2 f}{(d X)^2 }\right|_{\ov{X}}\partial^a \bar{\psi}\partial^a \bar{\psi}>0, \quad a=1,2,3\label{pre-Lor-cond}
\end{eqnarray}
which using \eqref{choice of psi} imply
\begin{equation}
\left. \frac{df}{d X }\right|_{\ov{X}}+\frac{\alpha^2}{2}\left. \frac{d^2 f}{(d X)^2 }\right|_{\ov{X}}<0, \qquad  \left. \frac{df}{d X }\right|_{\ov{X}}>0.
\label{Lor-cond}
\end{equation}
Note that, due to the choice of a solution of the form \eqref{choice of psi}, the conditions \eqref{pre-Lor-cond} are not only implying a pseudo-Riemannian signature but also the constancy of the metric components, which hence can be easily rescaled so to take the familiar Minkowskian form ${\rm diag}(-1,+1,+1,+1)$.

Of course, there are many possible choices of $f(X)$ and $\alpha$ which can fulfill the above requirements.
For example, we can pick up the specific combination
\begin{equation}\label{solution for f}
f(X)= -X^2+ X,  \qquad \frac{1}{3}<\alpha^2<\frac{1}{2}.
\end{equation}
However, in what follows we should not make use of any particular form of $f(X)$ and $\alpha$ and simply assume that they are such that \eqref{Lor-cond} are satisfied.

To summarize, since $g^\mn _k\equiv \eta^\mn$, $\forall k$, the (free) perturbations $\varphi_i$ are propagating on a Minkowski space, even though the fundamental theory  is Euclidean (\cf \eqref{fundamental Lagrangian}).
At this point few remarks are in order.

So far, our theory does not posses any fundamental speed scale. This is natural since the fundamental theory is Euclidean. At this level, there is no coordinate with time dimension and therefore one cannot define a constant with speed dimension. The invariant speed $c$, which will relate the length $x_0$ to an actual time parameter $t$, could be determined experimentally by first introducing a coordinate with time dimension (as it would be natural to do given the hyperbolic form of the equations of motion for the perturbations)  and then by defining $c$ as the signal speed associated to light cones in the effective spacetime\footnote{
Noticeably, a similar situation is encountered in the von Ignatowsky derivation of Special Relativity \cite{ignatowsky} where, given a list of simple axioms, one derives the existence of a universal speed, observer independent, which is not fixed a priori to be the speed of light but has to be identified via actual experiments.}.

Second, a comment is due about our choice of the background solution around which we have considered the dynamics for perturbations. It is obvious that within our model this choice is arbitrary. It simply shows that there are some background solutions $\bar{\psi}$ for which a pseudo-Riemannian metric can emerge. Obviously, different background solutions could lead to alternative metrics, \eg one could also obtain the Euclidean metric $\delta_\mn$ (for example if $\psi$ is constant), a degenerate metric or more complicated structures according to the possible solutions $\psi$. While it is conceivable that in a more complicate model we could have some mechanism for selecting the specific background solution that leads to an emergent Lorentzian signature, it is not obvious at all that such a feature should be built in the emergent theory. In fact, one generally minimizes an energy functional to select the ground state of the theory. However, when looking at Lorentzian signature emergence starting from an Euclidean set up as in our model, there is no initial notion of time and hence no energy functional to minimize. It is therefore unclear how a ground state could be selected from within the emergent system. 

 On the other hand, it is also conceivable that the actual background solution in which the initial system of fields (\ref{fundamental Lagrangian}) emerges from the fundamental (pre-manifold) theory, can be
 depending on
the conditions for which the ``condensation" of the fundamental objects takes place. In this sense,  the right ground state or background solution would be selected from minimizing some functional defined  at the level of the atoms of space-time. To use an analogy, the same fundamental constituents, \eg carbon atoms, can form very different materials, diamond or graphite, depending on the external conditions during the process of formation. Similarly, in a Bose--Einstein condensation the characteristics of the background solution (the classical wave function of the condensate), such as density and phase, are determined by physical elements (like the shape of the EM trap or the number and kind of atoms involved) which pre-exist the formation of the condensate.

In conclusion, we have identified the fundamental Lagrangian so that the  perturbations $\varphi_i$ have a kinematic term determined by the Minkowski metric.
\begin{equation}\label{Lagrangian for perturbations}
\lll_{\rm eff}(\varphi_1,...\varphi_N )= \sum_i  \eta^{\mn} \partial_\mu\varphi_i\partial_\nu\varphi_i.
\end{equation}
In this sense, we have a toy-model for the emergence of the Poincar\'e symmetries.
This construction can be seen as a generalization of the typical situation in analogue models of gravity \cite{LRR} where one has Poincar\'e symmetries emerging from fundamental Galilean symmetries \cite{LRR}. However, let us stress that in our case no preferred system of reference is present in the underling field theory given that the fundamental Lagrangian is endowed with a full Euclidean group $ISO(4)$.
Moreover, the emergence of a pseudo-Riemannian metric is in our model free of the usual problems encountered in the context of continuous signature change (\eg degenerate metrics)\footnote{ Bose-Einstein condensate analogue models of signature change events have indeed been considered in the literature together with the associated particle production (see \eg \cite{sign-chg} and references therein). These works are rather different from the one presented here as in that context the fundamental Lagrangian is non-relativistic and simply the emergent metric for the perturbations can have Lorentzian or Riemannian signature depending on the experimental possibility of changing at will the sign of the atomic interaction. } given that the former arises as a feature of the dynamics of perturbations around some solution of the equations of motion. Similarly one can see that  the invariance under Lorentz transformations is only an approximate property of the field equations (as usual for emergent systems), valid up to some order in perturbation theory.  In particular, if we analyze the third order contribution in \eqref{expansion} we get\footnote{We are in the mono-metric case, so that $F(X_1,..., X_N)=f(X_1)+...+f(X_N)$, and  $\psi_k=\psi$, $\forall k$.}
 \begin{equation}\label{cubic contribution}
\partial_\alpha \varphi_k \partial_\beta \varphi_k\partial_\gamma \varphi_k \left(\left. \frac{d^2 f}{(d X_k)^2}\right|_{\ov{X}}\partial^\alpha\psi \delta^{\beta\gamma}+ \frac{1}{6}\left. \frac{d^3 f}{(d X_k)^3}\right|_{\ov{X}}(\partial^\alpha \psi \partial^\beta \psi \partial^\gamma \psi ) \right).
\end{equation}
This contribution is clearly not Lorentz invariant if the solution $\psi$ pinpoints a specific direction,  as for example when the Minkowski metric is emergent. As a matter of fact our theory will show {\ae}ther like effects beyond second order.

So far, we have hence generalized and extended results familiar to the analogue gravity community. However, as said, a typical drawback of analogue gravity models is related to the fact that they show only the emergence of a background Lorentzian geometry while they are unable to reproduce a geometrodynamics of any sort. In what follows, we shall show that our model overcomes this drawback and indeed is able to describe the emergence of a  theory for scalar gravity. This theory will come out to be the only known other theory of gravitation, apart from General Relativity, which satisfy the strong equivalence principle~\cite{Will}, \ie Nordstr\"om gravity (for details see the Appendix).

\section{Emergence of Nordstr\"om gravity}

In this section, we describe how we can recover a relativistic scalar gravity theory from a Lagrangian of the type \eqref{fundamental Lagrangian}, when ground state is such that  the perturbations are living (at the lowest order in perturbation theory) in a Minkowski spacetime.  So, let us start from the truncated Lagrangian for the perturbations \eqref{Lagrangian for perturbations} that we obtained in the previous section. This Lagrangian can simply be  rewritten in terms of the (real) multiplet $\varphi= (\varphi_1,...,\varphi_N)$ as
\be \label{Lagrangian for perturbations 1}
\lll_{\rm eff}(\varphi)= \eta^\mn (\partial_\mu  \varphi)^T (\partial_\nu  \varphi).
\ee
This system has a global $O(N)$ symmetry which has emerged as well from the initial Lagrangian \eqref{fundamental Lagrangian}. It is hence quite natural to rewrite  the multiplet $\varphi$  by
introducing an amplitude characterized by a scalar field $\Phi(x)$  and a multiplet $\phi(x)$ with $N$ components such that\footnote{Our field redefinition is the generalization of the so-called Madelung representation \cite{LRR}.}
\begin{equation} \label{field redefinition}
\left(\begin{array}{c}\varphi_1\\\vdots
\\\varphi_N\end{array}\right)= \Phi\left(\begin{array}{c}\phi_1\\\vdots
\\\phi_N\end{array}\right),  \qquad \textrm{with } |\phi|^2\equiv\sum_i \phi_i^2=\ell^2.
\end{equation}
$\ell$ is an arbitrary length parameter to keep the dimension right. In particular, $\Phi$ is dimensionless and $\phi$ has the dimension of a length. $\Phi$ is the field invariant under $O(N)$ transformations, whereas  $\phi$ does transform under $O(N)$. As we shall see, this  field redefinition will provide us the means to identify gravity and matter degrees of freedom. The Lagrangian for the perturbations \eqref{Lagrangian for perturbations 1} reads now as\footnote{We use the normalization condition $|\phi|^2=\ell^2$, which implies in particular $\sum_i\phi_i \partial_\mu \phi_i=0$.}
\begin{equation}\label{new Lagrangian}
\lll_{\rm eff}(\varphi_1,...\varphi_N )\dr \lll_{\rm eff} (\Phi,\phi_1,...\phi_N)=\ell^2 \eta^{\mn} \partial_\mu\Phi\partial_\nu\Phi + \sum_i\Phi^2 \eta^{\mn}\partial_\mu\phi_i\partial_\nu\phi_i + \lambda(|\phi|^2-\ell^2),
\end{equation}
where $\lambda$ is a Lagrange multiplier. We recognize  in particular  the action for a non-linear sigma model given in terms of the fields $\phi_i$. The associated equations of motion are
\begin{eqnarray}
&& \eta^{\mn} (\ell^{2}\partial_\mu\partial_\nu\Phi - \Phi \sum_i\partial_\mu\phi_i\partial_\nu\phi_i) =0, \label{gravity EOM}\\
&& \eta^{\mn} (2\partial_\mu\Phi\partial_\nu\phi_i + \Phi^2 \partial_\mu\partial_\nu\phi_i+ \frac{1}{\ell^{2}} \partial_{\mu}\phi_{j}\partial_{\nu}\phi_{k}\delta^{jk} \phi_{i}) =0, \label{matter EOM}\\
&& |\phi|^2-\ell^2=0. \label{constraint}
\end{eqnarray}
If we introduce the (conformally flat) metric
\be\label{conformal-metric 1}
g_{\mn}(x)= \Phi^2(x)\eta_{\mn},
\ee
the equations of motion \eqref{matter EOM} can be simply rewritten as
\begin{equation}\label{matter}
(\sqrt{-g})\mone\partial_\mu(\sqrt{-g} g^{\mn} \partial_\nu \phi_i)+ \frac{1}{\ell^{2}} g^{\mu\nu} \partial_{\mu}\phi_{j}\partial_{\nu}\phi_{k}\delta^{jk} \phi_{i}= \Box_g \phi_i +\frac{1}{\ell^{2}} g^{\mu\nu} \partial_{\mu}\phi_{j}\partial_{\nu}\phi_{k}\delta^{jk} \phi_{i}=0,
\end{equation}
where we have introduced the d'Alembertian $\Box_g$ for the metric $g$ and used that $ \sqrt{-g}=\Phi^4$ and $g^{\mn}= \Phi^{-2}\eta^{\mn}$. (Incidentally, let us note that equation \eqref{matter EOM}  can be rewritten in the form \eqref{matter} using the metric redefinition \eqref{conformal-metric 1} only in four dimensions.)
To be consistent, the change of variable  $\Phi\dr g_\mn$ should be completed with the constraint that $g_\mn$ is conformally flat, that is
\begin{equation}\label{weyl tensor}
C_{\alpha\beta\gamma\delta}(g)=0,
\end{equation}
where $C_{\alpha\beta\gamma\delta}$ is the Weyl tensor.

Eq.~\eqref{matter} suggests  that \emph{the gravitational degree of freedom should be encoded in the scalar field} $\Phi$, whereas \emph{matter should be encoded in the} $\phi_i$. We are therefore aiming at a scalar theory of gravity with actions:
\begin{eqnarray}
S_{\rm eff}=\int dx^4\sqrt{-\eta}\, \lll_{\rm eff} =S_{\rm grav}+S_{\rm matter},\label{tot-action}\\
S_{\rm grav}= \ell^2 \int dx^4 \sqrt{-\eta}\, \eta^{\mn} \partial_\mu\Phi\partial_\nu\Phi ,\label{grav-action}\\
S_{\rm matter}=\int dx^4 \sqrt{-\eta}\,\left( \sum_i\Phi^2 \eta^{\mn}\partial_\mu\phi_i\partial_\nu\phi_i + \lambda(|\phi|^2-\ell^2)\right),\label{matt-action}
\end{eqnarray}
where we have explicitly written the volume element $\sqrt{-\eta}=1$ so to make clear that these actions are given in flat spacetime.

It is easy to see that the very same actions can be recast in the form of actions in a curved spacetime endowed with the metric \eqref{conformal-metric 1}. In particular for the matter action in \eqref{matt-action} one has
\begin{equation}\label{new Lagrangian for matter}
S_{\rm matter}= \int dx^4 \left( \sum_i \Phi^2 \eta^{\mn} \partial_\mu\phi_i\partial_\nu\phi_i + \lambda(|\phi|^2-\ell^2)\right) = \int \sqrt{-g}dx^4\left(\sum_ig^{\mn}\partial_\mu\phi_i\partial_\nu\phi_i + \lambda'(|\phi|^2-\ell^2)\right),\nn
\end{equation}
where we  have suitably rescaled the Lagrange multiplier to $\lambda'$. This allows to construct the stress-energy tensor $T_\mn$ for the non-linear sigma model, and its trace $\textbf{T}$ with respect to the metric $g$:
\begin{equation}
T_\mn=\frac{2}{\sqrt{-g}}\frac{\delta S_{\rm matter}}{\delta g^\mn}= \sum_i \left (\partial_\mu\phi_i\partial_\nu\phi_i - \demi g_\mn (g^{\alpha\beta}\partial_\alpha\phi_i\partial_\beta\phi_i) \right), \qquad \textbf{T}= g^\mn T_\mn = - \Phi^{-2} \sum_i \eta^{\mn}\partial_\mu\phi_i\partial_\nu\phi_i. \nn
\end{equation}
Finally, the above result, together with the recognition that the Ricci scalar $\textbf{R}$, associated to the metric $g_{\mn}$, can be written as $\textbf{R}=  - 6 {\Box_\eta \Phi}/{\Phi^{3}}$, allows us to rewrite Eq.~\eqref{gravity EOM} as the Einstein--Fokker equation\footnote{The very same equation could be derived from the action \eqref{tot-action} if one also notices that the gravitational action \eqref{grav-action} can be rewritten as well in a curved spacetime form by a simple integration by parts and the addition of a set of Lagrange multipliers implementing the vanishing of the Weyl tensor.}
\begin{equation}\label{einstein fokker}
\Box_\eta \Phi = \frac{1}{\ell^2}\eta^{\mn}\Phi \sum_i\partial_\mu\phi_i\partial_\nu\phi_i \quad \Leftrightarrow\quad  \textbf{R } = \frac{6}{\ell^2}  \textbf{T}.
\end{equation}

In summary, we can gather together the  equations of motion \eqref{matter}, \eqref{weyl tensor},   \eqref{einstein fokker}, obtained by introducing the metric \eqref{conformal-metric 1}, we have
\begin{eqnarray}
&& \textbf{R } =   \frac{6}{\ell^2}  \textbf{T}, \qquad C_{\alpha\beta\gamma\delta}=0. \label{emergent1}\\
&& \Box_g \phi_i + \frac{1}{\ell^{2}} g^{\mu\nu} \partial_{\mu}\phi_{j}\partial_{\nu}\phi_{k}\delta^{jk} \phi_{i}=0, \qquad |\phi|^2-\ell^2=0. \label{emergent3}
\end{eqnarray}
We recognize the equations of motion as those for Nordstr\"om gravity
\begin{equation}\label{Nord}
\textbf{R}= 24  \pi G_{\rm N}\, \textbf{T}, \qquad C_{\alpha\beta\gamma\delta}=0,
\end{equation}
coupled to a non-linear sigma model. Indeed, the rewriting of \eqref{gravity EOM}-\eqref{constraint} into the form \eqref{emergent1}-\eqref{emergent3}, is a special case of the procedure suggested by Einstein and Fokker so to cast Nordstr\"om gravity in a geometrical form \cite{fokker}.

We see from the above equation that the Newton constant $G_{\rm N}$ in our model has to be proportional to $\ell^{-2}$. However, in identifying the exact relation between the two quantities, some care has to be given to the fact that the stress-energy tensors appearing respectively in equation \eqref{emergent1} and equation \eqref{Nord} do not share the same dimensions. This is  due to the fact that the fields $\phi_i$ have the dimension of a length rather than the usual one of an energy. This implies that in order to really compare the expressions one has to suitably rescale our fields with a dimensional factor, $\Xi$, which in the end  would combine with $\ell$ so to produce an energy, ${\rm dim}[\ell\,\Xi]={\rm energy}$. In particular,  is easy to check that one has to assume ${4\pi} \ell^2 \Xi^2 \equiv  E^2_{\rm Planck}$ in order to recover the standard value of $G_{\rm N}$ (assuming $c$ as the observed speed of signals and $\hbar$ as the quantum of action).  As a final remark, we should stress that the scale $\ell$ is completely arbitrary within the emergent system and in principle should be derived from the physics of the ``atoms of spacetime" whose large N limit gives rise to \eqref{fundamental Lagrangian}.

Accidentally,  the above discussion also shows that, once the fields are suitably rescaled so to have the right dimensions,  the constraint appearing in Eq.\eqref{emergent3} is fixing the norm of the multiplet to be equal to the square of the Planck energy. This implies that the interaction terms in the aforementioned equation are indeed Planck-suppressed and hence negligible at low energy.
This should not be a surprise, given that in the end $\ell\Xi$ is the only energy scale present in our model. It is conceivable that more complicate frameworks, possibly endowed with many dimensional constants, will introduce a hierarchy of energy scales and hence break the degeneracy between the scale of gravity and the scale of matter interactions.

\section{Remarks and conclusions}
In the first section, we have considered fields that live in a Euclidean space, and showed that there exists a class of Lagrangians such that the perturbations around some classical solutions $\bar{\psi}$ propagate in a Minkowski spacetime. In this case $\bar{\psi}$ is essentially picking up a preferred direction, so that we have a spontaneous symmetry breaking of the Euclidean symmetry. The apparent change of signature is free of the problems usually met in signature change frameworks since the theory is fundamentally Euclidean. Lorentz symmetry is only approximate, and in this sense it is \emph{emergent}.

The main lesson we want to emphasize here is that \emph{Lorentzian signature can emerge from a fundamental Euclidean theory} and this process can in principle be  reconstructed by observers living in the emergent system. In fact, while from the perturbations 
point of view it is a priori difficult to see the fundamental Euclidean nature of the world, this could be guessed from the fact that some Lorentz symmetry breaking would appear at high energy (in our case in the form of a non-dynamical ether field).

In the second part, using a natural field redefinition, we have identified from the perturbations $\varphi_i$, a scalar field $\Phi$  encoding gravitational degrees of freedom and a set of scalar fields $\phi_i$ (a non-linear sigma model) encoding matter. In this sense, gravity and matter are both emergent at the same level\footnote{ Note that while the purely gravitational sector of Nordstr\"om gravity could be reproduced by a single free massless scalar field, the emergence of gravity {\em and} matter requires of course the introduction of several independent degrees of freedom, \ie many fields.}. This approach is then rather different from the one of analogue models of gravity where one usually identifies the analogue of the gravitational  degrees of freedom with the ``background" fields, \ie the condensate or the solution $\psi$ of the equations of motion. Indeed, following this line of thought in looking for a theory of gravitational dynamics, we would be led to require that the fundamental field theory \eqref{fundamental Lagrangian} must be endowed with diffeomorphisms invariance from the very start  --- the symmetries of the background are identical by construction to the ones of the fundamental theory. This would imply that one would have  to obtain gravity from a theory which is already diffeomorphisms invariant and hence most probably with a form very close to some known theory of gravitation. For these reasons, we do expect that if an emergent picture is indeed appropriate for gravitation, then it should be of the sort presented here, with both matter and gravity emerging at the same level\footnote{Of course, we cannot exclude that a full fledged theory of gravity could emerge,  together with the notion of manifold,  in a single step from the eventual semiclassical/large number limit of the fundamental objects. In this case, however, we would still have a very different picture from the one envisaged in analogue models of gravity.}.

In particular, this allows not only for an emergent local Lorentz invariance for the perturbations dynamics but it leads as well to an emergent diffeomorphisms invariance. In fact, we saw how the equations of motion \eqref{gravity EOM} and \eqref{matter EOM}  could be  rewritten in a completely equivalent way using a conformally flat metric \eqref{conformal-metric 1}.  Most noticeably, they can be rewritten in an evidently diffeomorphisms invariant form, from the point of view of ``matter fields observers"\footnote{Following the standard hole argument (see \eg \cite{sonego}), this also implies that the coordinates $x_\mu$, used to parameterized our theory, do not have any physical meaning from the point of view of the $\phi_i$ ``matter observers". They are merely parameters.}. In agreement with the fact that diffeomorphisms invariance is emergent in our system,  it can be noted that the cubic contribution \eqref{cubic contribution} ends up breaking it at the same level it breaks Lorentz invariance. Moreover, our derivation obviously holds for small perturbations $\varphi_i$, and hence small $\Phi$, implying that in our framework one would predict strong deviations from the weak field limit of the theory whenever the gravitational field becomes very large. We think this is an intriguing aspect of this proposal which might deserve further investigation.

Coming back to the emergence of diffeomerphism invariance, we note that Nordstr\"om gravity is also a nice framework for discussing the subtle distinction between background independence and diffeomorphisms invariance \cite{bckgd}. We call background some geometrical degrees of freedom that are not dynamical. For example, in General Relativity the topology of the manifold and its dimension, or the signature of the metric, can be considered as (trivial) background quantities. We can therefore have some specific background structures while still having diffeomorphisms invariance. Nordstr\"om gravity is encoded in conformally flat metrics.
If one considers fields which are conformally coupled to the metric (such as the electromagnetic field), these fields only see the metric $\eta_\mn$ which is of course not dynamical. The Minkowski metric can be see then as a background structure, this is what one may call a ``prior geometry" (\eg see \cite{MTW}).
One may hence say that diffeomorphism invariance is somewhat of a weaker form in Nordstr\"om gravity with respect the one present in general relativity.

In particular, while the essence of diffeomorphism invariance in GR is encoded in the associated Hamiltonian constraints, these are not defined in the present formulation of Nordstr\"om gravity. Furthermore,  in the most general implementations of Norstr\"om theory, quantities can be built which manifestly include the background structure $\eta_{\mn}$ and hence are not diffeomorphism invariant.  
However, within our model, the prior geometry cannot be detected 
\footnote{We want to thank S. Sonego for discussions on this point.}.  
Indeed, in order to detect the Minkowski background, one should be  
able to propose a method to pinpoint the conformal factor $\Phi^2$ in  
the relation $g_{\mu\nu}=\Phi^2\eta_{\mu\nu}$. However, a careful  
analysis shows that this is actually impossible. Let us elaborate on  
this point.
If we perform a conformal transformation, $x^{\mu}\rightarrow  
\bar{x}^{\mu}(x)$, the equations of motions associated to  
\eqref{Lagrangian for perturbations 1} are transforming like
\begin{equation}
  \Box_{\eta}\varphi_i = 0 \rightarrow \Box_{\bar{\eta}}\varphi_i=0,
\end{equation}
where $\eta$ and $\bar{\eta}$ are two different Minkowski metrics  
related by some conformal factor $\lambda(x)$. Therefore, $\eta$ and $ 
\bar{\eta}$ are indistinguishable, due to conformal invariance the  
equations of motion for $\varphi_{i}$. Hence, what appears to be a  
background structure, namely $\eta_{\mu\nu}$, is ambiguously defined,  
and the coordinates $x^{\mu}$ in which the equations of motion for the  
fields $\varphi_{i}$ are written have no operational meaning, they are  
mere labels. Furthermore, this ambiguity in the definition of what  
would be called a background structure implies an ambiguity on the  
definition of the conformal factor relating the physical metric to the  
would-be background structure. In this sense, within this very  
specific implementation of the model which has conformal invariance,  
there is no Minkowski geometry as a background. There is a background  
structure, which is the conformal structure of Minkowski spacetime.  
This is a mild limitation of our simple toy model as a diffeomorphism  
invariant, background independent system.

Of course, the above discussion holds only at the lowest order in the  
fields $\varphi_{i}$. As previously discussed, higher orders in  
perturbation theory will produce terms like \eqref {cubic  
contribution} producing a breaking of the conformal symmetry and hence  
the appearance of the background structures, {\it i.e.} the Euclidean  
space and the $\partial_{\mu}\bar{\psi}$ which have selected the  
timelike direction.

Finally, Nordstr\"om gravity is only a scalar gravity theory, which has been  falsified by experiments (\eg the theory does not predict the bending of light).   In order to obtain a more physical theory, in particular General Relativity, one should surely look for more complicated emergent Lagrangians than \eqref{fundamental Lagrangian}. Of course, one would in this case aim to obtain the emergence of a theory characterized by spin-2 gravitons (while in Nordstr\"om theory the graviton is just a scalar). This would open a door to a possible conflict with the so called Weinberg-Witten theorem \cite{WW}. However, there are many ways in which such a theorem can be evaded (see \eg \cite{Jenkins}) and in particular one may guess that analogue models inspired mechanisms like the one discussed here will generically lead to Lagrangian which show Lorentz and diffeomorphism invariance only as approximate symmetries for the lowest order in the perturbative expansion.

It is unclear which sort of generalization may still lead to some viable gravitational theory from the perturbations dynamics. For example, the simple addition of a potential will in general prevent the selection of a preferred direction, except in regions where the potential is almost flat. Moreover, it would also spoil the metric interpretation of the theory. For example, the terms $|\varphi|^n$ for $n\geq1$ and $\neq 4$ cannot be rewritten as an interaction between the matter field fields $\phi$ living on the conformal metric $\Phi^2\eta_\mn$, when using the change of variables \eqref{field redefinition} (although it is interesting to note that  a $|\varphi|^4$ term would give  Nordstr\"om gravity with a cosmological constant).
However, this ``rigidity" of the model is most probably due to its simplicity:  considering a more complex emergent field theory with fields such as spinors or tensors could possibly allow to have a preferred direction pinpointed while giving rise to more physical Lagrangians for the perturbations. We see this as a possible development of the research presented in this work.

\appendix
\section{Nordstr\"om gravity}\label{NG}
Nordstr\"om gravity was a key step in the search for a relativistic theory of gravity \cite{nordstrom,norton,Giulini:2006ry}. It is a scalar gravity theory  which is a natural generalization of Newtonian gravity.  
Indeed after the inception of Special Relativity in 1905, it was quickly realized that a relativistic theory for gravity was needed. The naive extension consisted into generalizing the Poisson equation to a relativistic equation,
$$\Delta \Phi = \rho \dr \square \Phi = \rho,$$ but some ambiguities both for the Newton equation (the dynamics of matter in a gravitational field) and for the (relativistic) source for gravity, arose.   In particular, this theory was not accounting for the characteristic non-linear behaviour of gravity which should be expected by the special relativistic equivalence of energy and mass.

After several attempts\footnote{See \cite{norton} for all references and some very interesting historical summary of the different steps that prepared the advent of General Relativity.}, Nordstr\"om  managed to propose an improved theory
using the newly introduced stress-energy tensor as a source for gravity\footnote{In Nordstr\"om's (second) formulation, the relativistic alter-ego of the Poisson equation is $\Phi\Box_\eta\Phi= \textbf{T}_\eta$, where $\textbf{T}_\eta= T_\mn \eta^\mn$ and $T_\mn $ is the matter stress-energy tensor defined with respect to the Minkowski metric. This formulation is not in accordance to the modern formulation encoding gravitational degrees of freedom in the metric.}.
Einstein and Fokker then showed that Nordstr\"om's theory could be encoded in purely geometry manner \cite{fokker} using the metric formalism. In the modern language, Nordstr\"om gravity (minimally coupled to a scalar field) can be summarized as follows
\begin{eqnarray}
C_{\alpha\beta\gamma\delta}=0,\label{EOM2} \\
\textbf{R}= \ka \textbf{T}, \label{EOM1}\\
 (\square_g + m^2)\phi =0,\label{EOM3}
\end{eqnarray}
where $\textbf{T}$ is the trace of the stress energy tensor, defined with respect to $g_\mn$, for the matter field $\phi$, $\textbf{R}$ is the Ricci scalar for $g_\mn$ and $\ka$ is proportional to the Newton constant $G_{\rm N}$ (indeed $\kappa=24\pi G_{\rm N}$).

Equation \eqref{EOM2} encodes the fact that the Weyl tensor is zero, \ie that the metric $g_\mn$ is conformally flat and there is some coordinate system in which it takes the form
\begin{equation}\label{conformal-metric}
g_\mn = \Phi^2 \eta_\mn,
\end{equation}
where $\eta_\mn$ is the Minkowski metric.
Equation \eqref{EOM1} is \emph{not} the trace of the Einstein equations restricted to the conformal metric, due to the absence of a minus sign in front of the Ricci scalar.
Equation \eqref{EOM3} is  simply the equation of motion of the scalar field $\phi$ propagating in the metric $g_\mn$.

This theory was introduced before  Eddington's observation showing the bending of light. It is clear that since the metric is conformally flat, no bending of light can occur there, so that Nordstr\"om gravity is experimentally ruled out. Note however that a massless minimally coupled scalar field    would see non trivial gravitational effects. Nordstr\"om theory predicts also an anomalous perihelion precession of Mercury, which disagrees in both sign and magnitude with the observed anomalous precession.

Nordstr\"om gravity is  interesting as it shares many aspects of General Relativity, such as diffeomorphisms invariance and the strong equivalence principle~\cite{Will}. It is also an example of the subtle distinction between diffeomorphisms invariant theories and background invariant theories as we discussed in the Discussion section. In general, scalar gravities have attracted much attention due to their simplicity and similitude with General Relativity (see \eg  \cite{misner} which also contains  a complete list of references for scalar gravity theories).

\acknowledgements{
The authors wish to thank S. Sonego, T. Sotirou, M. Visser and S. Weinfurtner for useful remarks and critical readings of the manuscript.
}


\begin{thebibliography}{99}

\bibitem{Bardeen:1973gs}
  J.~M.~Bardeen, B.~Carter and S.~W.~Hawking,
  \emph{The Four laws of black hole mechanics},
  Commun.\ Math.\ Phys.\  {\bf 31}, 161 (1973).

\bibitem{Hawking:1974sw}
  S.~W.~Hawking,
  \emph{Particle Creation By Black Holes},
  Commun.\ Math.\ Phys.\  {\bf 43}, 199 (1975)
  [Erratum-ibid.\  {\bf 46}, 206 (1976)].

\bibitem{Jacobson}
  T.~Jacobson,
  \emph{Thermodynamics of space-time: The Einstein equation of state},
  Phys.\ Rev.\ Lett.\  {\bf 75}, 1260 (1995)
  \href{http://arxiv.org/abs/gr-qc/9504004}{[arXiv:gr-qc/9504004]};\\
  C.~Eling, R.~Guedens and T.~Jacobson,
  \emph{Non-equilibrium Thermodynamics of Spacetime},
  Phys.\ Rev.\ Lett.\  {\bf 96}, 121301 (2006)
  \href{http://arxiv.org/abs/gr-qc/0602001}{[arXiv:gr-qc/0602001]}.
\bibitem{Isham}
C.~J.~Isham, {\em``Structural issues in quantum gravity''},
\href{http://arxiv.org/abs/gr-qc/9510063}{[arXiv:gr-qc/9510063]};
\\
J.~Butterfield and C.~J.~Isham, {\em ``On the emergence of time in
quantum gravity''}, \href{http://arxiv.org/abs/gr-qc/9901024}{[arXiv:gr-qc/9901024]}.

\bibitem{Kuchar}
K.~V.~Kuchar, {\em ``Time And Interpretations Of Quantum
Gravity''}, In ``Winnipeg 1991, Proceedings, General relativity
and relativistic astrophysics", 211-314.

\bibitem{Barbour}
  J.~Barbour,
 \emph{ Dynamics of pure shape, relativity and the problem of time},
  Lect.\ Notes Phys.\  {\bf 633}, 15 (2003)
  \href{http://arxiv.org/abs/gr-qc/0309089}{[arXiv:gr-qc/0309089]}.

\bibitem{BLH}
  B.~L.~Hu,
  \emph{Can spacetime be a condensate?},
  Int.\ J.\ Theor.\ Phys.\  {\bf 44}, 1785 (2005)
   \href{http://arxiv.org/abs/gr-qc/0503067}{[arXiv:gr-qc/0503067]}.

\bibitem{Bombelli:1987aa}
  L.~Bombelli, J.~H.~Lee, D.~Meyer and R.~Sorkin,
\emph{ Space-Time As A Causal Set},
  Phys.\ Rev.\ Lett.\  {\bf 59}, 521 (1987).

\bibitem{Oriti:2007qd}
  D.~Oriti,
  \emph{Group field theory as the microscopic description of the quantum spacetime
  fluid: a new perspective on the continuum in quantum gravity,}
  \href{http://arxiv.org/abs/gr-qc/0710.3276}{arXiv:0710.3276 [gr-qc]}.

\bibitem{Konopka:2008hp}
    T.~Konopka, \emph{ Statistical Mechanics of Graphity Models},
  \href{http://arxiv.org/abs/gr-qc/0805.2283}{arXiv:0805.2283 [hep-th]};
  T.~Konopka, F.~Markopoulou and S.~Severini,
  \emph{Quantum Graphity: a model of emergent locality},
  Phys.\ Rev.\  D {\bf 77}, 104029 (2008)
  \href{http://arxiv.org/abs/hep-th/0801.0861}{arXiv:0801.0861 [hep-th]};
  T.~Konopka, F.~Markopoulou and L.~Smolin,
 \emph{Quantum graphity},
   \href{http://arxiv.org/abs/hep-th/0801.0861}{arXiv:hep-th/0611197}.


\bibitem{Dreyer}
  O.~Dreyer,
  \emph{Why things fall},
    Proceedings of From Quantum to Emergent Gravity: Theory and Phenomenology, Trieste, Italy, 11-15 Jun 2007.
   \href{http://arxiv.org/abs/gr-qc/0710.4350}{arXiv:0710.4350 [gr-qc]};
   O.~Dreyer,
  \emph{Emergent general relativity}.
  Contribution to book ``Towards Quantum Gravity". Edited by D. Oriti. Cambridge University Press, 2006.
  \href{http://arxiv.org/abs/gr-qc/0604075}{[arXiv:gr-qc/0604075]}.

\bibitem{LRR}   C.~Barcelo, S.~Liberati and M.~Visser,
  \emph{Analogue gravity},
  Living Rev.\ Rel.\  {\bf 8} (2005) 12
  \href{http://arxiv.org/abs/gr-qc/0505065}{[arXiv:gr-qc/0505065]}.



\bibitem{L-Vol}
  G.~E.~Volovik,
 \emph{Vacuum energy: myths and reality},
  Int.\ J.\ Mod.\ Phys.\  D {\bf 15}, 1987 (2006)
 \href{http://arxiv.org/abs/gr-qc/0604062}{[arXiv:gr-qc/0604062]}.

 \bibitem{greensite}
  J.~Greensite,
\emph{Dynamical origin of the Lorentzian signature of space-time},
Phys.\ Lett.\  B {\bf 300} (1993) 34
 \href{http://arxiv.org/abs/gr-qc/9210008}{[arXiv:gr-qc/9210008]}.

\bibitem{Ellis:2006sq}
 G.~F.~R.~Ellis,
 \emph{Physics in the real universe: Time and spacetime},
 Gen.\ Rel.\ Grav.\  {\bf 38} (2006) 1797
 \href{http://arxiv.org/abs/gr-qc/0605049}{ [arXiv:gr-qc/0605049]}.

\bibitem{ignatowsky}   S.~Liberati, S.~Sonego and M.~Visser,
\emph{Faster-than-c signals, special relativity, and causality},
  Annals Phys.\  {\bf 298} (2002) 167
   \href{http://arxiv.org/abs/gr-qc/0107091}{[arXiv:gr-qc/0107091]}.
   W.A.~von Ignatowsky,
\emph{Einige allgemeine Bemerkungen zum Relativit\"atsprinzip},
Verh. Deutsch. Phys. Ges. 12, 788-796 (1910);
\emph{Einige allgemeine Bemerkungen zum Relativit\"atsprinzip},
Phys. Zeitsch. 11, 972-976
(1910);
\emph{Das Relativit\"atsprinzip},
Arch. Math. Phys. 3 (17), 1-24; (18), 17-41 (1911);
\emph{Eine Bemerkung zu meiner Arbeit 'Einige allgemeine Bemerkungen zum Relativit¨atsprinzip},
Phys. Zeitsch. 12, 779 (1911).

\bibitem{sign-chg}
  S.~Weinfurtner, A.~White and M.~Visser,
  \emph{Trans-Planckian physics and signature change events in Bose gas
  hydrodynamics},
  Phys.\ Rev.\  D {\bf 76}, 124008 (2007)
  \href{http://arxiv.org/abs/gr-qc/0703117}{[arXiv:gr-qc/0703117]}.


\bibitem{Will}
  C.~M.~Will,
  \emph{The confrontation between general relativity and experiment},
  Living Rev.\ Rel.\  {\bf 9}, 3 (2005)
  \href{http://www.livingreviews.org/lrr-2006-3}{http://www.livingreviews.org/lrr-2006-3};
  \href{http://arxiv.org/abs/gr-qc/0510072}{[arXiv:gr-qc/0510072]};\\
 J.-M.\ G\'erard,
\emph{The strong equivalence principle from a gravitational gauge structure},
Classical and Quantum Gravity {\bf 24} (7), 1867--1877 (2007).
 \href{http://arxiv.org/abs/gr-qc/0607019} {[arXiv:gr-qc/0607019]}

\bibitem{MTW}  C.~W.~Misner, K.~S.~Thorne and J.~A.~Wheeler,
\emph{ Gravitation},
 San Francisco 1973

\bibitem{sonego}    H.~Westman and S.~Sonego,
  \emph{Coordinates, observables and symmetry in relativity},
  \href{http://arxiv.org/abs/gr-qc/0711.2651}{arXiv:0711.2651 [gr-qc]}.


\bibitem{bckgd}
  D.~Giulini,
  \emph{Some remarks on the notions of general covariance and background
independence}, Lect.\ Notes Phys.\  {\bf 721} (2007) 105
  \href{http://arxiv.org/abs/gr-qc/0603087}{[arXiv:gr-qc/0603087]}.


\bibitem{WW}
  S.~Weinberg and E.~Witten,
  \emph{Limits On Massless Particles},
  Phys.\ Lett.\  B {\bf 96}, 59 (1980).

\bibitem{Jenkins}
  A.~Jenkins,
  \emph{Topics in particle physics and cosmology beyond the standard model},
  \href{http://arxiv.org/abs/hep-th/0607239}{arXiv:hep-th/0607239}.

\bibitem{nordstrom} G.~Nordstr\"om,
\emph{Zur Theorie der Gravitation vom Standpunkt des Relativitätsprinzip},
Annalen der Physik,
42: 533-554.

\bibitem{norton} J. Norton {\it Einstein, Nordstr\"om and the early Demise of Lorentz-covariant, Scalar Theories of Gravitation}, Archive for History of Exact Sciences, 45 (1992),  \href{http://www.pitt.edu/~jdnorton/papers/Nordstroem.pdf}{http://www.pitt.edu/~jdnorton/papers/Nordstroem.pdf}. J. Norton, {\it Einstein and Nordstr\"om: Some Lesser Known Thought Experiments in Gravitation},  The Attraction of Gravitation: New Studies in History of General Relativity. Boston: Birkh\"auser.  \href{http://www.pitt.edu/~jdnorton/papers/einstein-nordstroem-HGR3.pdf}{http://www.pitt.edu/~jdnorton/papers/einstein-nordstroem-HGR3.pdf}

\bibitem{Giulini:2006ry}
  D.~Giulini,
  \emph{What is (not) wrong with scalar gravity?}
  \href{http://arxiv.org/abs/gr-qc/0611100}{[arXiv:gr-qc/0611100]}.

\bibitem{fokker}
  A.~Einstein and A.~D.~Fokker,
  Nordstr\"om's Theory of Gravitation from the Point of View of the Absolute Differential Calculus,
  Annalen Phys.\  {\bf 44}, 321 (1914)
  [Annalen Phys.\  {\bf 14}, 500 (2005)].

\bibitem{misner} K.~Watt and C.~W.~Misner,
\emph{Relativistic scalar gravity: A laboratory for numerical relativity},
 \href{http://arxiv.org/abs/gr-qc/9910032}{arXiv:gr-qc/9910032};
   W.~T.~Ni,
\emph{Theoretical frameworks for testing relativistic gravity IV:  a
compendium of
 metric theories of gravity and their post-newtonian limits},
  Astroph. Journal {\bf 176}:769 (1972)





\end{thebibliography}
\end{document}